# Plasma Screening Effects in Stark Broadening: A Fully Relativistic Close-Coupling Approach

Chao Wu,[1] Yong Wu,[2,3] Yu Hao Zhu,[4] Ming Li,[1] Jian Guo Wang,[2] and Xiang Gao [2, *]

1. College of Science, Xi'an University of Science and Technology, Xi'an 710054, China

2. National Key Laboratory of Computational Physics, Institute of Applied Physics and Computational Mathematics, P. O. Box 8009, Beijing 100088, China

3. HEDPS, Center for Applied Physics and Technology, Peking University, Beijing 100084, China

4. College of Science, Xi'an University of Architecture and Technology, Xi'an 710055, China

Stark broadening of spectral lines in plasmas is a cornerstone of opacity modeling and plasma diagnostics, with critical implications for controlled fusion and astrophysics. Despite recent advances in fully quantum-mechanical close-coupling calculations for electron-impact broadening, the impact of denser plasma environments remains largely unexplored due to theoretical bottlenecks associated with electron-ion collision processes. Based on our newly developed close-coupling theory for electron-ion collisions in plasmas, which resolves the problem of extracting short-range scattering phase shifts, we introduce a fully relativistic close-coupling approach for the Stark broadening that incorporates plasma screening effects. Systematic investigations of hydrogenic radiators reveal distinct patterns of line broadening dependence on plasma conditions, offering valuable insights for plasma diagnostic applications. Furthermore, we provide a quantum-mechanical interpretation of the screening factor commonly introduced in semi-classical impact theories. This work establishes a robust foundation for future studies on complex atomic systems in high-density plasmas.

## I. INTRODUCTION

Stark broadening is the pressure broadening of the spectral lines of a radiating atom (ion) caused by charged particles in the surrounding plasma [1]. It provides an essential input for opacity modeling [2]. Moreover, an accurate understanding of Stark broadening is crucial for the spectroscopic diagnostics of laboratory and astrophysical plasmas [3]. Therefore, it has extensive applications in astrophysics, controlled fusion, and high-energy-density physics research [4]. Previous experimental and theoretical studies have mostly focused on low-density plasmas [5]. Extending the theory of impact broadening of spectral lines to denser plasma environments is of great significance for both fundamental research and related applied studies.

Spectral lines from atoms and ions embedded in plasmas are shaped not only by the individual atomic properties of the radiator but also by the statistical influence of the surrounding plasma environment. Theoretical description of Stark broadening is therefore very challenging, as it lies at the interface of atomic physics, plasma physics, and statistical physics [6,7]. Although the theoretical framework for a unified treatment of ions and electrons in Stark broadening has been proposed based on the large-scale molecular dynamics simulations [8] as well as Green's function approaches [7,9], the large number of physical ingredients involved makes such a comprehensive description impractical for most applications. Consequently, due to the large mass difference between ions and electrons, most theoretical investigations of Stark broadening consider the contributions from ions and electrons separately according to the ratio between the characteristic collision velocity $v$ and impact parameter $b$ [1].

Ion perturbers move slowly and typically interact with radiators at relatively large distances, so that the ratio $v/b$ is small. As a consequence, the electric field produced by the ions varies only weakly over the radiative timescale of the transition. Under these conditions, many ions can substantially influence a radiator simultaneously, and a single binary-collision picture is not appropriate. This physical situation justifies the quasi-static approximation, in which the atomic response is described by the Stark effect in an external field whose instantaneous value is determined by the collective contribution of many ions, thereby giving rise to a statistical description of the plasma microfield. The pioneering work of Holtsmark established the foundation for this approach, which remains a cornerstone of ion broadening theory [10,11].

Electrons, in contrast to ions, are fast and highly penetrating perturbers. Their interactions with radiators are dominated by close encounters involving small impact parameters, so the ratio $v/b$ is large. The electron contribution is usually much larger than the ion contribution [1]. There are three primary methods for describing the electron contribution to Stark broadening, i.e., the impact theory [12-16], the relaxation theory [6,17-19], and the kinetic theory [20]. The general forms of the line shape expressions from these three theories are the same, with the differences entering mainly through the specific form of the electron-impact broadening operator [1]. Although the relaxation theory and the kinetic theory are mathematically more rigorous and provide statistical descriptions of electron-impact broadening by accounting for the cumulative effect of many electron collisions on the radiating states, many approximations involved in the impact theory also have to be made for their practical use, such as the binary collision assumption [18]. In this framework, line broadening arises from a sequence of short-lived, well-separated binary collisions, and the effect of an individual collision is characterized by a scattering matrix connecting the radiator states before and after the encounter. Averaging over many independent collisions then yields an electron-impact broadening operator that determines the spectral width and shift. Therefore, the key physical quantity for electron broadening is the modeling of scattering processes, which are closely related to electron-atom (ion) collision theory, as recognized in early works by Regemorter [21] and Sahal-Bréchot [22].

The treatments of the scattering processes for electron-radiator collisions can be broadly divided into two categories. When the plasma density is low, the collision can be treated as a weak, short-duration perturbation, and one can use the semi-classical approaches, which are computationally efficient and have been widely used in practical Stark broadening calculations [1,22-24]. In these approaches, the internal structure of the radiator is treated quantum mechanically, while the perturbing electron is assumed to move along a classical trajectory, typically approximated as a straight line or hyperbolic path. On the other hand, as the plasma density increases, close encounters induce strong short-range interactions, and the probability of electron penetration into the inner atomic region drastically increases, where a full quantum-mechanical description of both the electron and the radiator within a unified scattering framework is required [12-14,25-30]. There are two common approximations for the quantum scattering processes: perturbative first-order Born/distorted-wave or second-order approximation for the electron-impact broadening operator [1,25,26], and the dipole approximation for the Coulomb interaction between atoms and plasma particles [1]. Only very recently have full-order close-coupling scattering calculations without the dipole approximation been developed [11,17,18,27-30], and the importance of resonances in electron collisions has been demonstrated [27,29,30].

In denser plasma environments, the increased probability of electron penetration necessitates not only fully quantum-mechanical scattering calculations but also a fully relativistic treatment. In addition, many-body collision effects are expected to become significant, among which the collective screening effect is the most prominent, where the collective response strongly modifies interparticle interactions. As a result, individual charged particles no longer interact through the bare Coulomb potential, but rather through an effective interaction screened by the surrounding plasma, thereby affecting the energy-level structure and dynamical processes [31-36]. For electron-collision processes, particle interactions in plasmas are commonly described using screened potentials, such as the Debye-Hückel potential [37] or atomic-state-dependent effective potentials [38], which represent the statistical averaging of the collective effects from the surrounding plasma's ion interactions. Most existing quantum-mechanical calculations [25-30] are performed for isolated atoms (ions) and do not systematically incorporate plasma screening, not only because of the scarcity of experimental data on line broadening in dense plasmas, but also due to intrinsic difficulties in collision theory, particularly for ionic systems. Because screening alters the long-range Coulomb boundary conditions, it complicates the accurate extraction of screened short-range scattering phase shifts, making it difficult to obtain reliable scattering matrices for Stark broadening calculations. It should be noted that in some semi-classical treatments, screening is introduced through a cutoff, typically to truncate the range of the Coulomb interaction (or, equivalently, the maximum impact parameter) for avoiding the divergence of classical-trajectory integrals, while its detailed physical picture has yet to be clarified [1].

Recently, we have developed a fully relativistic close-coupling theoretical approach for electron-impact broadening, which solves the multichannel scattering problem without invoking a weak-interaction assumption [27-30]. This approach provides a non-perturbative description of the collision dynamics, allowing close encounters, channel coupling, exchange effects, and resonance structures to be treated consistently. Although computationally more demanding, such approaches offer a more complete and systematic description of electron-radiator collisions, particularly in regimes dominated by low-energy electrons and strong interactions. We have also developed a close-coupling theory that incorporates plasma screening by means of an R-matrix method for determining scattering matrices under screened interaction potentials [39].

## II. THEORETICAL METHOD

### A. Stark line profiles

In this work, we further extend our previous electron-impact broadening code [29,30] to include contributions from both ions and electrons. The line profile is obtained by averaging over the ionic microfield distribution, while electron collisions enter through the electron-impact broadening operator $\phi$. The Stark line profile of one-electron systems is given by [1,11]

$$L(\Delta\omega) = -\left(\frac{1}{\pi}\right) \text{Re Tr} \int_0^\infty dF \, W(F)\{\Delta_d[i\Delta\omega - iCF + \phi]^{-1}\}, \quad (1)$$

where $\Delta\omega$ is the frequency separation from the unperturbed line, $\Delta_d$ is the dipole operator, $W(F)$ is the Griem-Holtsmark microfield distribution [1], and $C$ is the linear Stark coefficient [1],

$$C = \frac{3}{2}n(n_1 - n_2)\frac{ea_0}{\hbar Z}, \quad (2)$$

where $n_1$, $n_2$ are parabolic quantum numbers. The $iCF$ term corresponds to the ion contribution. The real and imaginary parts of $\phi$ correspond to the electron-impact width $w$ and shift $d$, respectively, which are evaluated using the effective collision strengths obtained from quantum scattering calculations as [25,26]

$$\phi = w + id = 2N_e \left(\frac{\hbar}{m_e}\right)^2 \left(\frac{2m_e\pi}{k_B T_e}\right)^{1/2} \times \int_0^\infty \Gamma(\varepsilon) \times \exp\left(\frac{-\varepsilon}{k_B T_e}\right) d\left(\frac{\varepsilon}{k_B T_e}\right). \quad (3)$$

Here $T_e$, $N_e$, and $\varepsilon$ denote the electron temperature, electron density, and incident electron energy, respectively. The effective collision strength $\Gamma$ is obtained from the scattering matrices connecting the initial and final states of the radiative transition [29,30].

$$\Gamma = \sum_{J_i^T J_f^T j j' l l'} (-1)^{J_i + J_{i'} + 2J_f^T + j + j'} \frac{1}{2}(2J_i^T + 1)(2J_f^T + 1) \times \begin{Bmatrix} J_f^T & J_i^T & 1 \\ J_i & J_f & j \end{Bmatrix} \begin{Bmatrix} J_f^T & J_i^T & 1 \\ J_{i'} & J_{f'} & j' \end{Bmatrix}$$

$$\times \left[ \delta_{ll'}\delta_{jj'}\delta_{J_i J_{i'}}\delta_{J_f J_{f'}} - \left(S_I(J_{i'}l'j'J_i^T; J_i l j J_i^T) \times S_F^*(J_{f'}l'j'J_f^T; J_f l j J_f^T)\right) \right], \quad (4)$$

where $S_I$ and $S_F$ are the scattering matrices of the upper and lower states, respectively, and they are calculated under the same incident electron energy $\varepsilon$ of colliding electrons.

### B. Relativistic R-matrix determination of screened scattering matrices

To incorporate plasma screening, we replace the electron-ion as well as electron-electron Coulomb interactions with Debye-screened potentials in the electron scattering calculations. More specifically, the electron-atom (ion) scattering problem is treated using a fully relativistic close-coupling R-matrix method, in which the complete electron-electron interaction is included without invoking the dipole approximation, following the formulation developed in Ref. [39]. The long-range polarization effects in the outer region are treated rigorously, without adopting the approximations commonly employed in earlier works [40]. Within this framework, the screened short-range scattering matrix is determined by matching the asymptotic boundary condition with the R-matrix $R_{ii'}$ at the matching point $r_a$, i.e.,

$$F_{ij}(r_a) = \sum_{i'} R_{ii'} \left[ r_a \frac{dF_{i'j}(r)}{dr} - bF_{i'j}(r) \right]_{r=r_a}, \tag{5}$$

with $F_{i'j}$ and $dF_{i'j}(r)/dr$ the channel radial wavefunction and its derivative, respectively. $b$ is a constant parameter in the R-matrix definition which usually takes zero. At large distances, where the potential is dominated by the centrifugal term, $F_{ij}$ can be expressed as:

$$F_{ij}(r) = S_i(kr)\delta_{ij} + C_i(kr)K_{ij}, \tag{6}$$

where $K_{ij}$ is the scattering $K$ matrix and the scattering $S$ matrix in Eq. (4) is calculated by the definition $S = (1 - iK)^{-1}(1 + iK)$. Unlike the usual electron-atom scattering calculations, where $S_i(kr)$ and $C_i(kr)$ are the regular and irregular Riccati-Bessel functions (or Coulomb wavefunctions for ion target), respectively, the most crucial step here is to use numerically calculated regular and irregular scattering wavefunctions $S_i(kr)$ and $C_i(kr)$ of the screened Coulomb potential, which have a phase difference of $\pi/2$. In Ref. [39], we have developed a numerical algorithm for this purpose which was also adopted in the present work. In this way, we can extract a well-defined short-range scattering matrix which is independent of the choice of asymptotic matching point $r_a$.

It should be emphasized that the effective $\Gamma$ defined in Eq. (4) differs fundamentally from the conventional collision strengths or scattering cross sections commonly used in electron-atom collision calculations. In conventional scattering calculations, the relevant quantities are inelastic cross sections from the specific ground or excited states. In this case, as the inelastic processes happen in a region close to the target, the short-range scattering phase shifts are insensitive to the choice of the $S_i(kr)$ and $C_i(kr)$ in the asymptotic region. Therefore, previous distorted-wave calculations using the pure Coulomb wavefunctions could give reasonable results [36,41-43]. On the other hand, for the present electron-impact broadening calculations, the effective $\Gamma$ is related to the total collision cross sections for excited atoms or ions. Therefore, by using the unitary property of the scattering matrix, the scattering matrix $S_I$ and $S_F$ in Eq. (4) are closely related to the elastic collision processes, which are very sensitive to the choice of the $S_i(kr)$ and $C_i(kr)$. Only the physically correct choice of the asymptotic wavefunctions can yield a well-defined short-range scattering matrix which is independent of the matching point, especially for the ion target.

### III. RESULTS AND DISCUSSION

Fig. 1(a) shows $\Gamma$ for the hydrogen 3d[+]-2p[+] component at several Debye screening lengths. As the screening becomes stronger (smaller $D$), $\Gamma$ decreases systematically over most of the incident energy range. An exception is observed at $D=30$ a.u., where $\Gamma$ rises noticeably relative to the $D=50$ a.u. case in the low-energy region. Under weak screening ($D=200$ a.u.), a comparison of Γ results with (solid line) and without (dotted line) the polarization effect demonstrates that this effect enhances Γ in the low-energy region. In contrast, under strong screening ($D=30$ a.u.), the near overlap of results with (solid line) and without (dotted line) the polarization effect across the entire energy range indicates that the long-range polarization contribution is significantly suppressed. This observation contrasts with the expected prominent effect under weaker screening, directly demonstrating the suppression mechanism as screening intensifies. Fig. 1(b) presents the corresponding Γ for the $3d^+$-$2p^+$ transition of $He^+$. With increasing screening, the resonance structures associated with higher Rydberg series are progressively damped, while resonances from lower series shift to higher energies; meanwhile, the overall magnitude of Γ also decreases as screening strengthens. In contrast, the Γ values (dotted line) obtained when the short-range scattering matrix is extracted using unscreened asymptotic Coulomb wavefunctions are noticeably larger than those from all properly screened treatments, highlighting the crucial role of short-range scattering matrix extraction. Such an approach, commonly adopted in earlier distorted-wave theories [36,41-43], can lead to unphysical Stark widths in high-density plasma conditions.

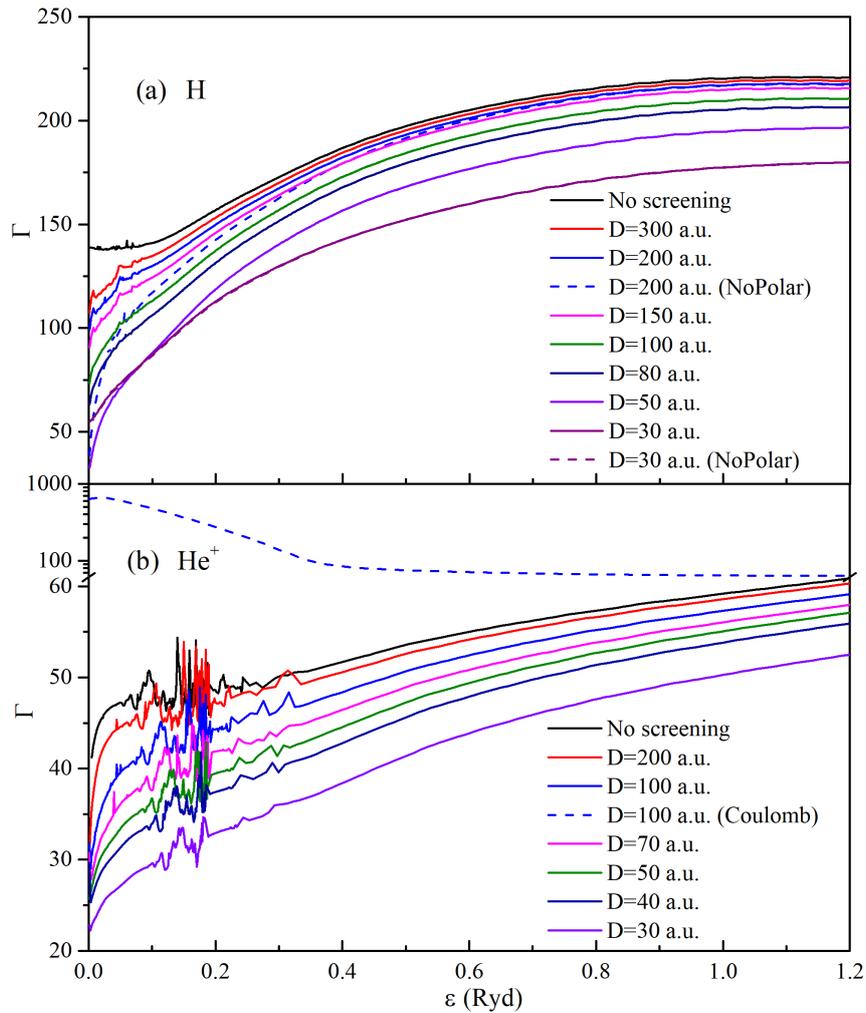

Fig. 1. Screening suppresses low-energy collision strengths Γ for the $3d^+$-$2p^+$ transition in H (a) and $He^+$ (b).

Relativistic quantum-mechanical calculations naturally give rise to fine-structure splittings of atomic levels. Under experimental conditions, however, the individual fine-structure components strongly overlap and are observed as a single spectral feature. Throughout this work, the reported Stark widths therefore correspond to oscillator-strength-weighted averages over all fine-structure components of a given transition.[28] To verify the correctness of our new code, Figs. 2 and 3 present benchmark comparisons under low-density plasma conditions. In Fig. 2, our calculated shifts of the $H_\beta$ line are compared with the experimental measurements of Wiese [44], as well as with results from relaxation theory [18] and kinetic theory [18]. The present results show good agreement with the measured values. By comparison, the shifts predicted by kinetic theory are several times larger than those obtained in the present calculations, possibly due to the limitations of the approximations employed in the specific calculations, as discussed in Ref. [18]. In contrast, the relaxation theory results are in close agreement with our values.

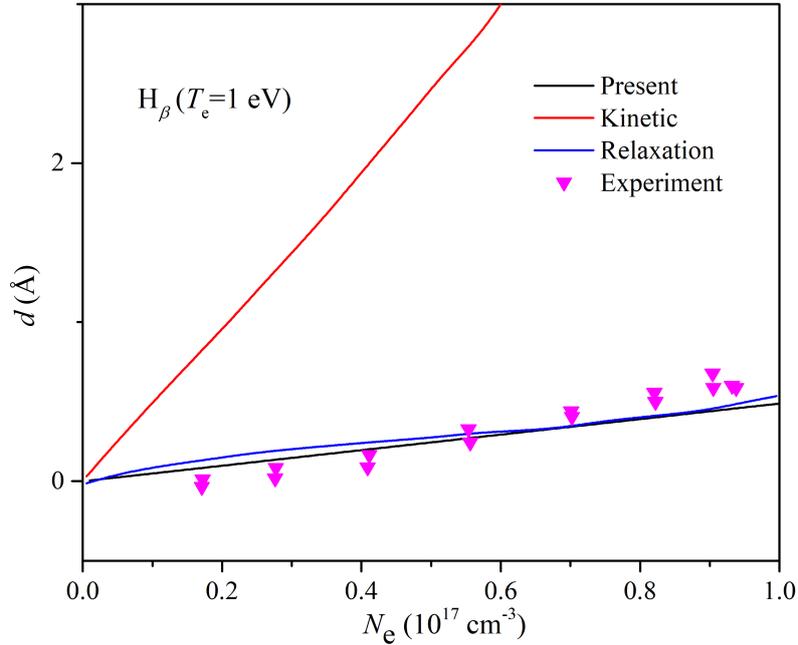

Fig. 2. Comparison of calculated $H_\beta$ line shift obtained from different theoretical models [18] and measured values [44].

In Fig. 3, our calculated $Ly_\alpha$ line profiles are compared with representative results from existing theoretical approaches, including semi-classical models (VCS) [17], classical particle simulations [45], and fully quantum-mechanical calculations [17]. Overall, our quantum-mechanical results are consistent with these established calculations, confirming the reliability of the present approach.

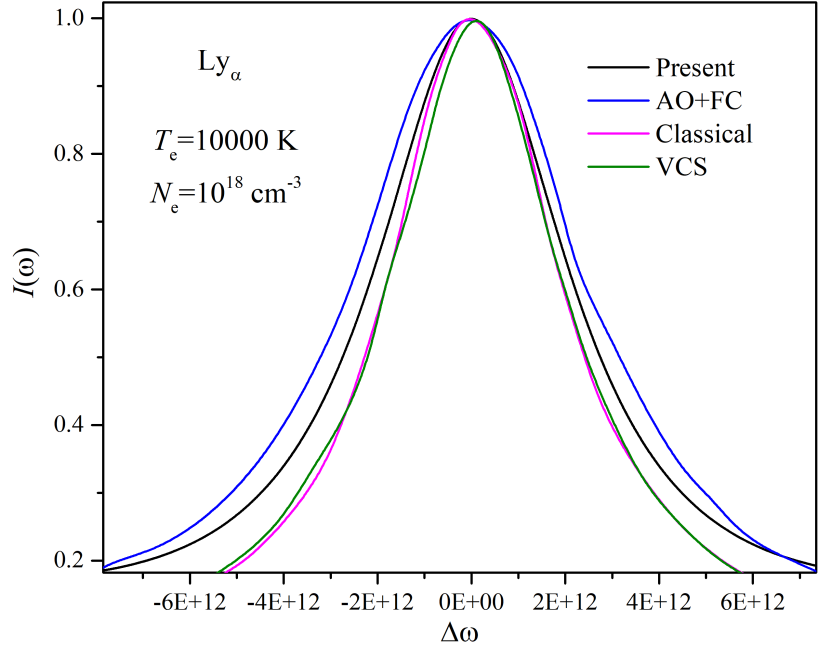

Fig. 3. Comparison of calculated Ly$_\alpha$ line profiles obtained from different theoretical models [17, 18, 45].

Figs. 4 and 5 examine the dependence of the H$_\alpha$ Stark width on electron density for neutral hydrogen and singly ionized helium, focusing on high-density regimes where plasma screening becomes significant. Fig. 4 compares our calculated H$_\alpha$ widths for hydrogen with high-density experimental measurements by John [46], which constitute a stringent benchmark. Semi-classical results by Griem [1] and Vidal [47] are also shown for comparison. Three quantum-mechanical calculations are presented: an isolated calculation with polarization (blue), a screened calculation without polarization (red), and a screened calculation with polarization (black). For the isolated case, as electron density increases, discrepancies between theory and experiment become more pronounced. At the highest measured density ($N_e=1.39\times10^{20}$ cm$^{-3}$), the isolated result overestimates the experimental width by a factor of 2.79. Plasma screening substantially reduces the line width, whereas the long-range polarization effect is significant only at low densities. At the lowest electron density considered ($N_e=8.63\times10^{17}$ cm$^{-3}$), the influence of screening can be ignored, whereas polarization increases the width by about 12%. As the density increases, the polarization effect is progressively suppressed and becomes insignificant. In contrast, including screening reduces the line width by up to 48% at densities around $10^{20}$ cm$^{-3}$, leading to markedly improved agreement with experiment. At this highest density point, the theory-to-experiment ratio drops to 1.44, and the average ratio across the studied density range is 1.33. The difference between the results of Vidal [47] and the experiment is significant, whereas Griem's calculations [1] reproduce the experimental density dependence well in the low-density region. However, the two semi-classical models exhibit trends that differ markedly from our results, raising questions about the reliability of their extrapolation to high-density plasma conditions.

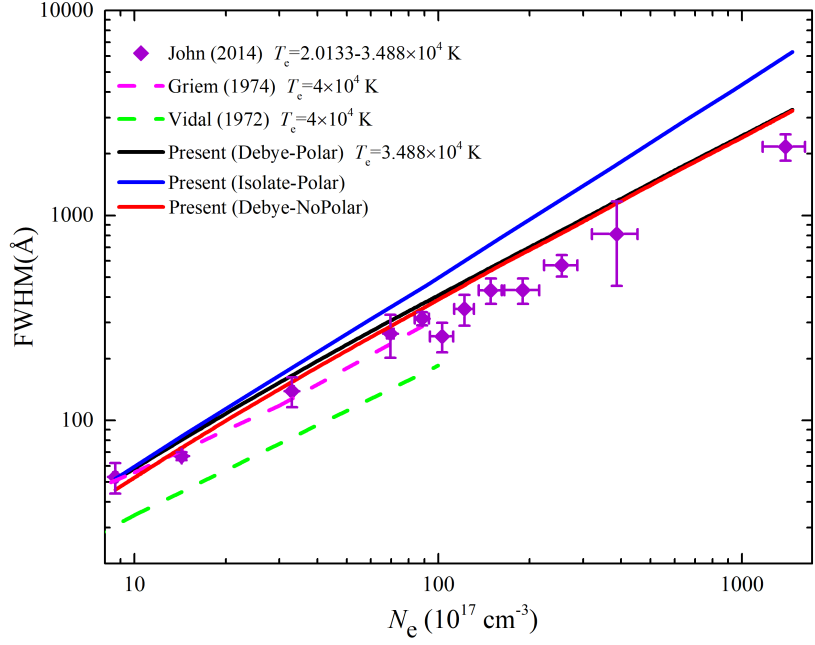

Fig. 4. Comparison of Stark broadening of the $H_\alpha$ line of H, semi-classical results from Griem [1] and Vidal [47], and experimental data from John [46].

Fig. 5 presents corresponding results for the $H_\alpha$ Stark widths of He$^+$. Besides the isolated and screened results (without/with polarization), a fourth calculation is also shown for comparison (cyan), where scattering phase shifts are extracted using Coulomb wavefunctions as asymptotic solutions in a screened plasma. In Debye plasmas, the use of unscreened Coulomb asymptotic functions yields systematically larger electron-impact widths than properly screened calculations, demonstrating that such treatments overestimate the widths. This indicates that accurate extraction of screened short-range phase shifts is essential for Stark width calculations of ionized radiators in plasmas. For He$^+$, the experimental densities [48] are relatively low. Consequently, over this measured range, the screened and isolated calculations remain close, with plasma screening exerting only a modest influence. The long-range polarization interaction affects the widths mainly at low densities; at the lowest density considered ($N_e$=8×10$^{17}$ cm$^{-3}$), where it increases the Stark width by about 8.5%. At the highest density ($N_e$=1×10$^{20}$ cm$^{-3}$), however, screening reduces the width by about 30%. Our screened results agree well with the measurements, giving an average theory-to-experiment ratio of 1.23. In addition, our screened results are also in good agreement with Griem's calculations [1]. The close agreement in both magnitude and trend between the semi-classical results and the screened calculations indicates that the short-range scattering matrix obtained from Eq. (6) provides the quantum-mechanical physical picture of the screening factor used in semi-classical theories, i.e., the screened short-range scattering processes are the key processes for electron-impact broadening.

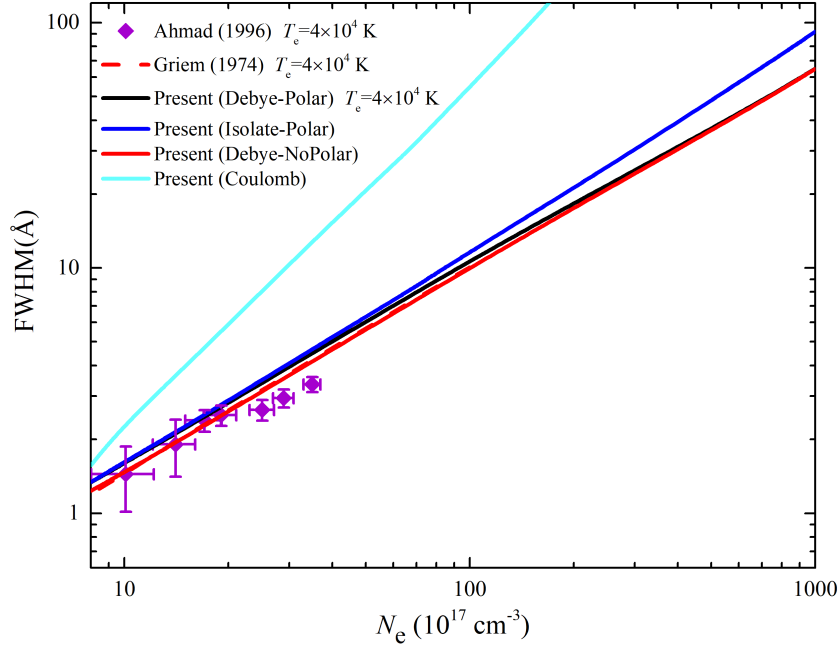

Fig. 5. Comparison of Stark broadening of the $H_\alpha$ line of $He^+$, semi-classical results from Griem [1], and experimental data from Ahmad [48].

Fig. 6 examines the dependence of the $H_\alpha$ Stark width on electron temperature for neutral hydrogen and singly ionized helium. For the isolated case, the Stark width increases monotonically as $T_e$ decreases. When plasma screening is included, this temperature dependence is substantially weakened. For H (Fig. 6a), the Stark width becomes nearly insensitive to $T_e$ at high electron densities and even shows a slight decrease on the lower $T_e$ side of the explored range. At $N_e=1\times10^{19}$ cm$^{-3}$ the turnover appears below $T_e\approx2.5\times10^4$ K, whereas at $N_e=1\times10^{20}$ cm$^{-3}$ a similar behavior sets in at $T_e\approx12.5\times10^4$ K. A similar behavior is observed for $He^+$ (Fig. 6b). At $N_e=1\times10^{20}$ cm$^{-3}$, the Stark width shows a clear non-monotonic temperature dependence, with the turnover occurring around $T_e\approx7.5\times10^4$ K.

Considering that Griem's semi-classical dataset [1] is limited, a simple extrapolation based on its trend yields a similar non-monotonic behavior but somewhat qualitatively different temperature dependences. Although Griem's semi-classical results in the relatively low temperature regions compare well with our calculations, these different temperature behaviors indicate the importance of fully quantum-mechanical treatments over extended temperature and density regions. Under the same plasma conditions, the screening-induced changes in $He^+$ widths are smaller than those in H, reflecting the stronger Coulomb attraction in ionized radiators that partially offsets the screening of the electron-radiator interaction. These results indicate that while $H_\alpha$ width provides a viable density diagnostic in high-density plasmas, its effectiveness for temperature diagnostics is reduced due to a significantly weakened width-temperature dependence.

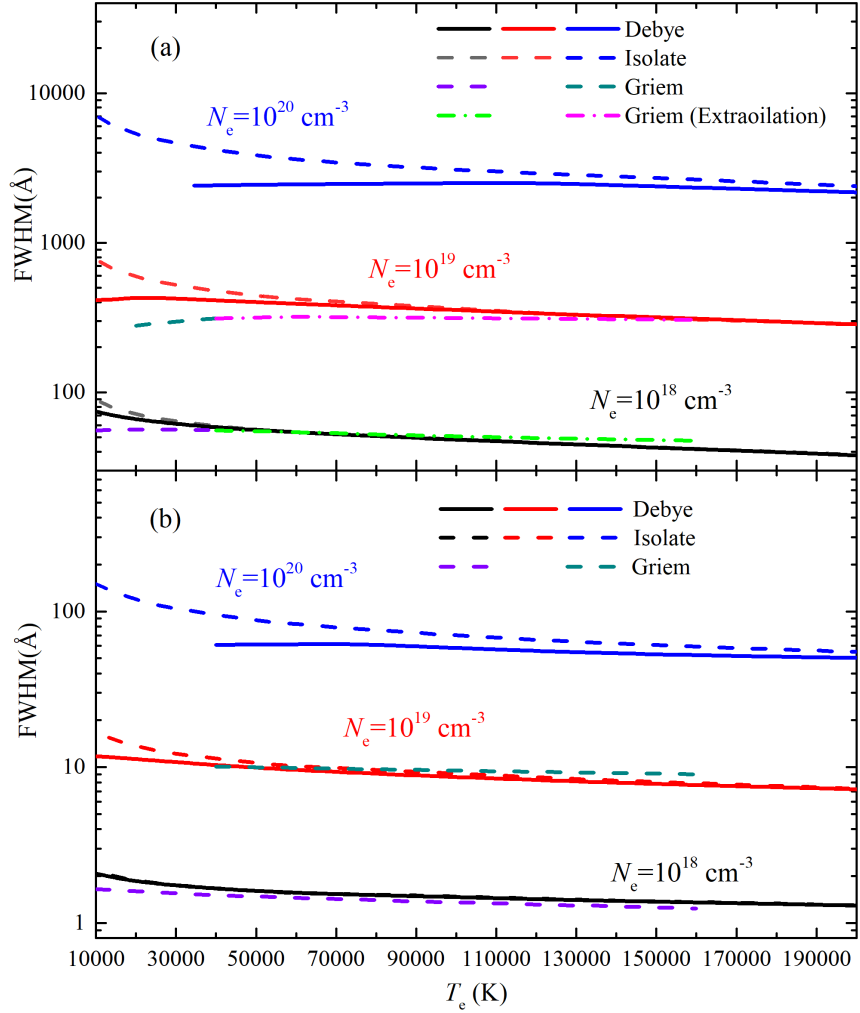

Fig. 6. Comparison of Stark broadening of the $H_\alpha$ line of H (a) and He$^+$ (b), semi-classical results from Griem [1].

## IV. CONCLUSION

In summary, by incorporating plasma screening directly at the level of electron-radiator scattering, we have established a quantum-mechanical framework for treating plasma-screening effects on Stark broadening in dense plasmas and applied it to hydrogenic lines. The resulting Stark widths exhibit significantly improved agreement with available experimental measurements at high electron densities. Our results demonstrate that plasma screening plays a key role in electron-impact broadening at high electron densities. In particular, for ionized radiators, we show that a proper treatment of screened boundary conditions is essential, as the use of unscreened Coulomb asymptotics leads to systematic overestimation of Stark widths. Comparisons with semi-classical theories indicates that short-range scattering phase shifts provide a quantum-mechanical physical picture for the screening factor used therein. The screening-induced suppression of low-energy collision strengths leads to a modified density dependence and, in some cases, a non-monotonic temperature dependence of the $H_\alpha$ width. This implies that $H_\alpha$ Stark broadening remains a robust diagnostic of electron density, whereas its sensitivity to temperature is significantly reduced in dense plasmas due to the screening-induced suppression of the width-temperature dependence. Although plasma screening is treated here within the Debye potential, this framework is readily applicable to general multielectron atomic systems. The present calculations provide a quantum-mechanical reference for assessing screening effects on Stark broadening. Extensions to more refined effective interaction models and ionic microfield distributions beyond the present Griem-Holtsmark treatment, as well as applications to other ionic radiators, are left for future work. In addition, further high-density experimental measurements would be particularly valuable for benchmarking plasma-screening effects.

## ACKNOWLEDGMENTS

This work is supported by the National Natural Science Foundation of China (Grant No. 12241410) and Shaanxi Province Natural Science Basic Research Program (No. 2025JC-YBQN-110).

[1] H. R. Griem, *Spectral line broadening by plasmas*, (Academic Press, New York, 1974).
[2] M. J. Seaton, Atomic data for opacity calculations XIII. Line profiles for transitions in hydrogenic ions, J. Phys. B: At. Mol. Opt. Phys. 23, 3255 (1990).
[3] R. E. Falcon, G. A. Rochau, J. E. Bailey et al., Laboratory measurements of white dwarf photospheric spectral lines: H$_\beta$, Astrophys. J. 806, 214 (2015).
[4] E. Oks, Review of recent advances in the analytical theory of stark broadening of hydrogenic spectral lines in plasmas: Applications to laboratory discharges and astrophysical objects, Atoms 6, 50 (2018).
[5] S. Djurović, B. Blagojević, and N. Konjević, Experimental and semiclassical stark widths and shifts for spectral lines of neutral and ionized atoms (a critical review of experimental and semiclassical data for the period 2008 through 2020), J. Phys. Chem. Ref. Data 52, 031503 (2023).
[6] U. Fano, Pressure broadening as a prototype of relaxation, Phys. Rev. 131, 259 (1963).
[7] L. Hitzchke, G. Ropke, T. Seifert et al., Green's function approach to the electron shift and broadening of spectral lines in non-ideal plasmas, J. Phys. B: At. Mol. Opt. Phys. 19, 2443 (1986).
[8] E. Stambulchik and C. A. Iglesias, Full radiator-perturber interaction in computer simulations of hydrogenic spectral line broadening by plasmas, High Energ. Dens. Phys. 6, 9 (2010).
[9] S. Alexiou, Green function approach to the theory of spectral line broadening in a plasma via the time-dependent microfield, Phys. Lett. A 143, 134 (1990).
[10] J. Holtsmark, Über die verbreiterung von spektrallinien, Ann. Phys. 58, 577 (1919).
[11] T. A. Gomez, T. Nagayama, P. B. Cho et al. Introduction to spectral line shape theory, J. Phys. B: At. Mol. Opt. Phys. 55, 034002 (2022).

[12] M. Baranger, General impact theory of pressure broadening, Phys. Rev. 112, 855 (1958).

[13] M. Baranger, Problem of overlapping lines in the theory of pressure broadening, Phys. Rev. 111, 494 (1958).

[14] M. Baranger, Simplified quantum-mechanical theory of pressure broadening, Phys. Rev. 111, 481 (1958).

[15] A. C. Kolb and H. R. Griem, Theory of line broadening in multiplet spectra, Phys. Rev. 111, 514 (1958).

[16] H. R. Griem, A. C. Kolb, and K. Y. Shen, Stark broadening of hydrogen lines in a plasma, Phys. Rev. 116, 4 (1959).

[17] T. A. Gomez, T. Nagayama, P. B. Cho et al., All-order full-Coulomb quantum spectral line-shape calculations, Phys. Rev. Lett. 127, 235001 (2021).

[18] T. A. Gomez, T. Nagayama, D. P. Kilcrease et al., Density-matrix correlations in the relaxation theory of electron broadening, Phys. Rev. A 98, 012505 (2018).

[19] E. W. Smith and C. F. Hooper, Relaxation theory of spectral line broadening in plasmas, Phys. Rev. 157, 126 (1967).

[20] T. Hussey, J. W. Dufty, and C. F. Hooper, Kinetic theory of spectral line broadening, Phys. Rev. A 12, 1084 (1975).

[21] V. H. Regemorter, Spectral line broadening, Annu. Rev. Astron. Astrophys. 3, 71 (1965).

[22] S. Sahal-Bréchot, Stark broadening of isolated lines in the impact approximation, Astron. Astrophys. 35, 319 (1974).

[23] M. S. Dimitrijević and S. Sahal-Bréchot, Stark broadening of neutral helium lines, J. Quant. Spectrosc. Radiat. Transfer 31, 301 (1984).

[24] H. R. Griem, M. Blaha, and P. C. Kepple, Stark-profile calculations for Lyman-series lines of one-electron ions in dense plasmas, Phys. Rev. A 19, 2421 (1979).

[25] H. Elabidi, N. B. Nessib, and S. Sahal-Bréchot, Quantum mechanical calculations of the electron-impact broadening of spectral lines for intermediate coupling, J. Phys. B: At. Mol. Opt. Phys. 37, 63 (2003).

[26] H. Elabidi, N. B. Nessib, M. Cornille et al., Electron impact broadening of spectral lines in Be-like ions: quantum calculations, J. Phys. B: At. Mol. Opt. Phys. 41, 025702 (2008).

[27] T. A. Gomez, T. Nagayama, C. J. Fontes et al., Effect of electron capture on spectral line broadening in hot dense plasmas, Phys. Rev. Lett. 124, 055003 (2020).

[28] B. Duan, M. A. Bari, Z. Q. Wu et al., Widths and shifts of spectral lines in He II ions produced by electron impact, Phys. Rev. A 86, 052502 (2012).

[29] C. Wu, X. Gao, Y. H. Zhu et al., Theoretical study of electron-impact broadening for highly charged Ar XV ion lines, Chinese Phys. B 32, 053101 (2023).

[30] C. Wu, X. Gao, Y. Wu et al., Relativistic quantum mechanical calculations of Stark broadening data of Li-like ions for stellar and laboratory plasma investigations, Mon. Not. R. Astron. Soc. 543, 3664 (2025).

[31] R. K. Janev, S. B. Zhang, and J. G. Wang, Review of quantum collision dynamics in Debye plasmas, Matter Radiat. Extrem. 1, 237 (2016).

[32] T. N. Chang, T. K. Fang, C. S. Wu et al., Atomic processes, including photo absorption, subject to outside charge-neutral plasma, Atoms 10, 16 (2022).

[33] T. N. Chang, T. K. Fang, C. S. Wu et al., Redshift of the isolated atomic emission line in dense plasma, Phys. Scripta 96, 124012 (2021).

[34] Y. Y. Qi, Y. Wu, and J. G. Wang, Photoionization of Li and radiative recombination of Li$^+$ in Debye plasmas, Phys. Plasmas 16, 033507 (2009).

[35] Y. Y. Qi, J. G. Wang, and R. K. Janev, Bound-bound transitions in hydrogen-like ions in dense quantum plasmas, Phys. Plasmas 23, 073302 (2016).

[36] Y. L. Ma, L. Liu, L. Y. Xie et al., Debye-screening effect on electron-impact excitation of helium-like Al$^{11+}$ and Fe$^{24+}$ ions, Chinese Phys. B 31, 043401 (2022).

[37] M. Das, B. K. Sahoo, and S. Pal, Plasma screening effects on the electronic structure of multiply charged Al ions using Debye and ion-sphere models, Phys. Rev. A 93, 052513 (2016).

[38] F. Y. Zhou, Y. Z. Qu, J. W. Gao et al., Atomic-state-dependent screening model for hot and warm dense plasmas, Commun. Phys. 4, 148 (2021).

[39] C. Wu, W. H. Xia, Y. Wu et al., R-matrix theory for electron-ion collisions in plasmas, arXiv:2601.19538 [physics.atom-ph].

[40] S. B. Zhang, J. G. Wang, and R. K. Janev, Electron-hydrogen-atom elastic and inelastic scattering with screened Coulomb interaction around the n=2 excitation threshold, Phys. Rev. A 81, 032707 (2010).

[41] B. L. Whitten, N. F. Lane, and J. C. Weisheit, Plasma-screening effects on electron-impact excitation of hydrogenic ions in dense plasmas, Phys. Rev. A 29. 945 (1984).

[42] Z. B. Chen, Electron-impact excitation of atoms or ions with the screened Coulomb potential, Phys. Plasmas 30, 032103 (2023).

[43] Y. D. Jung, Plasma-screening effects on the electron-impact excitation of hydrogenic ions in dense plasmas, Phys. Fluids B 5, 3432 (1993).

[44] W. L. Wiese, D. E. Kelleher, and D. R. Paquette, Detailed study of the Stark broadening of Balmer lines in a high-density plasma, Phys. Rev. A 6, 1132 (1972).

[45] P. B. Cho, T. A. Gomez, M. H. Montgomery et al., Simulation of Stark-broadened hydrogen Balmer-line shapes for DA white dwarf synthetic spectra, Astrophys. J. 927, 70 (2022).

[46] F. K. John and F. A. Nicole, Shift and width of the Balmer series Hα line at high electron density in a laser-produced plasma, J. Phys. B: At. Mol. Opt. Phys. 47, 155701 (2014).

[47] C. R. Vidal, J. Cooper, and E. W. Smith, Hydrogen Stark-broadening tables, Astrophys. J. Suppl. 25, 37 (1973).

[48] R. Ahmad, Measurement of the He II Balmer-α-line in a Z-pinch plasma, Nuovo Cimento D 18, 53 (1996).